\documentclass[aps, pra, twocolumn, superscriptaddress, showpacs, tightenlines, longbibliography]{revtex4-1}
\usepackage[colorlinks, breaklinks, citecolor=blue, linkcolor=blue,  urlcolor={blue}]{hyperref}
\usepackage{breakurl}
\usepackage{amsfonts}
\usepackage{amsmath}
\usepackage{amssymb}
\usepackage{mathrsfs}
\usepackage{epsfig}
\usepackage{graphicx}
\usepackage{bm}

\usepackage{url}
\usepackage[colorlinks]{hyperref}
\hypersetup{%
    plainpages=true,
    breaklinks=true,%not default in dvips mode, so we must specify
    hypertexnames=false,%not ideal, but needed when pagenums duplicate (`i' vs. `1')
    pageanchor=true,
    colorlinks=true,
    linkcolor={blue},
    citecolor={red},
    urlcolor={blue},
    anchorcolor={black}
}

\begin{document}

\title{Weak-force sensing with squeezed optomechanics}

\author{Wen Zhao}
\thanks{These authors contributed equally to this work.}
\affiliation{Key Laboratory of Low-Dimensional Quantum Structures and Quantum Control of Ministry of Education, Department of Physics and Synergetic Innovation Center for Quantum Effects and Applications, Hunan Normal University, Changsha 410081, China}
\affiliation{College of Physics and Materials Science, Henan Normal University, Xinxiang 453007, China}

\author{Sheng-Dian Zhang}
\thanks{These authors contributed equally to this work.}
\affiliation{Key Laboratory of Low-Dimensional Quantum Structures and Quantum Control of Ministry of Education, Department of Physics and Synergetic Innovation Center for Quantum Effects and Applications, Hunan Normal University, Changsha 410081, China}

\author{Adam Miranowicz}
\affiliation{Faculty of Physics, Adam Mickiewicz University,
    61-614 Pozna\'n, Poland}

\author{Hui Jing}
\email{jinghui73@foxmail.com}
\affiliation{Key Laboratory of Low-Dimensional Quantum Structures and Quantum Control of Ministry of Education, Department of Physics and Synergetic Innovation Center for Quantum Effects and Applications, Hunan Normal University, Changsha 410081, China}

\begin{abstract}
We investigate quantum-squeezing-enhanced weak-force sensing via a nonlinear optomechanical resonator containing a movable mechanical mirror and an optical parametric amplifier (OPA).
Herein, we determined that tuning the OPA parameters can considerably suppress quantum noise and substantially enhance force sensitivity, enabling the device to extensively surpass the standard quantum limit. This indicates that under realistic experimental conditions, we can achieve ultrahigh-precision quantum force sensing by harnessing nonlinear optomechanical devices.
\end{abstract}

%\ocis{(270.1670) Coherent optical effects; (120.4880) Optomechanics.}
\date{\today}

\maketitle

\makeatletter
\newcommand{\rmnum}[1]{\romannumeral #1}
\newcommand{\Rmnum}[1]{\expandafter\@slowromancap\romannumeral #1@}
\makeatother

\section{Introduction}

Cavity optomechanics (COM) has recently emerged as a versatile
platform for both fundamental studies of light-matter
interactions and practical applications ranging from optical
communication to quantum
metrology~\cite{clerk2010introduction,aspelmeyer2014cavity}.
In particular, COM sensors have achieved unprecedented sensitivity for measuring
mass~\cite{bin2019mass,liu2019realization},
acceleration~\cite{krause2012high,qvarfort2018gravimetry},
displacement~\cite{wilson2015measurement,
rossi2018measurement,matsumoto2019demonstration,Continuous2019Mason},
and force~\cite{caves1980measurement,
schreppler2014optically,zhou2018spectrometric,Force2019Ali,basiri2019precision,gebremariam2019enhancing}.
Moreover, the sensitivity of such sensors is constrained by the standard quantum limit
(SQL)~\cite{bowen2015quantum} or a lower
bound on the additional measurement uncertainty determined by the balance between
the shot and back-action noise.
However, the SQL has been surpassed using quantum non-demolition
techniques~\cite{braginskiui1975quantum,thorne1978quantum,braginsky1980quantum}
to achieve sub-SQL sensitivity via quantum entanglement~\cite{ma2017proposal,li2019sensitivity,carrasco2019fisher}
or squeezing~\cite{caves1981quantum,xu2014squeezing,
motazedifard2016force,clark2016observation,kampel2017improving,sudhir2017quantum,moller2017quantum,Otterpohl2019Squeezed}.
Squeezed states were studied early in 1927~\cite{Kennard1927}, although an explosion of interest in them was triggered 40 years ago when they were first used to detect gravitational waves
via supersensitive interferometry~\cite{Hollenhorst1979,caves1980measurement,Dodonov1980,caves1981quantum}.
Experiments by injecting squeezed light into a COM resonator have successfully demonstrated
sub-SQL sensitivity~\cite{xiao1987precision,aasi2013enhanced,hoff2013quantum,yap2018broadband}.
However, the inevitable injection losses hinder the
ultimate performance of COM sensors in practice.

To overcome this obstacle, COM sensing using an intracavity optical parametric amplifier (OPA) has recently been proposed to implement ultrahigh-precision
position detection~\cite{peano2015intracavity}.
This scheme has the advantage that all the information is imprinted
on the deamplified momentum quadrature,
which induces limited signal suppression
but simultaneously brings about a dramatic reduction in the
noise. Using such squeezed resonators, the SQL can be
attained precisely at a mechanical resonance without injection losses~\cite{peano2015intracavity}.
Additionally, sub-SQL sensitivity can be achieved
using an OPA in a dissipative COM system~\cite{huang2017robust}.
Nevertheless, none of these previous
works have considered the role of the optical phase,
particularly the OPA pump phase, in further enhancing the sensitivity.

This paper aims to fill this gap by discussing the effects of both the OPA gain
and pump phase on force sensing in a squeezed COM device. We show that tuning the
OPA parameters can considerably suppress quantum backaction noise
and enable the device to reach sub-SQL sensitivity at a smaller
COM coupling without losing quantum efficiency.
Unlike previous studies~\cite{peano2015intracavity}, our
study focuses on the squeezed quadrature and we determine that
in the limit where the cavity linewidth is much larger than any
measurement frequency of interest~\cite{wimmer2014coherent},
the squeezed momentum quadrature carries
complete information regarding the weak-force signals, and the information about the added force
contained in the position quadrature can be safely ignored.
Herein, we focus on canonical quadrature squeezing; however, other observables can also be squeezed,
such as the photon number~\cite{Huang2018pb,Li2019Nonreciprocal}.
Our research demonstrates that squeezed
COM devices are feasible and powerful enough to achieve ultrahigh
precision quantum measurement~\cite{taylor2013biological,Barzanjeh2015radar}.
\begin{figure*}[ht]
\centering
\includegraphics[width=6.3in]{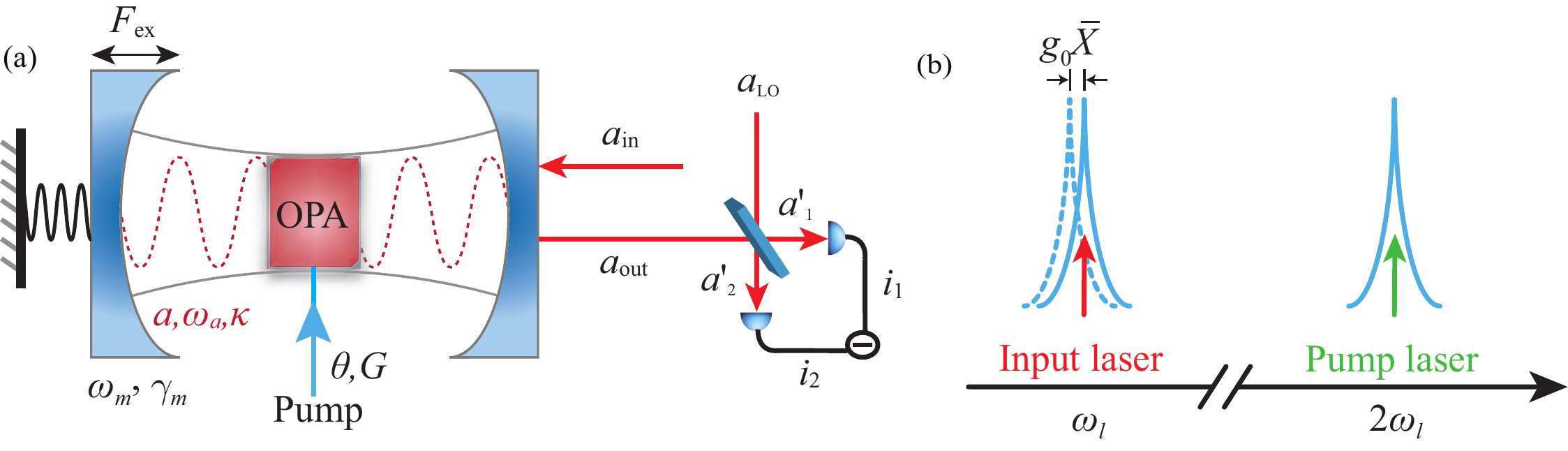}
\caption{(Color online) (a) Schematics of the cavity optomechanical (COM) system
consisting of a Fabry-P\'{e}rot cavity coupled to a degenerate
optical parametric amplifier (OPA). One or even two mirror
coatings are usually directly placed on the spherical and polished
surfaces of the nonlinear crystal that generates the squeezed
vacuum and produces parametric
amplification~\cite{wu1986generation,grangier1987squeezed}.
In this setup, an external weak force $F_\mathrm{ex}$ acting on
the mechanical resonator can be measured with homodyne detection.
(b) Frequency spectrum of the OPA-assisted COM system. An input
laser drives the optical cavity at frequency $\omega_l$, thus, a
pump laser beam with frequency $2\omega_l$ is applied to the
degenerate OPA. The dashed curve shows the static frequency shift
$g_0\bar{X}$ at the cavity resonance because of radiation
pressure.} \label{fig1}
\end{figure*}
%--------------------------------------------------------------------------------------%

\section{Model and solution}\label{Section 2}

Figure~\ref{fig1} shows the schematics of intracavity squeezing in
an OPA-assisted Fabry-P\'{e}rot cavity~\cite{wu1986generation,grangier1987squeezed}.
Previous researches on such
squeezed quantum systems have usually focused on enhanced mechanical
cooling~\cite{huang2009enhancement,huang2018improving,Ground2019Lau,Optomechanical2019Asjad} or
squeezing~\cite{squeezing2015Lyu,agarwal2016strong,gu2018enhanced} and enhanced light-matter coupling~\cite{Exponentially2018Qin,leroux2018enhancing,huang2009normal}. When light pressure couples to a movable mirror, coherent states are transformed into squeezed states of light~\cite{braginski1967ponderomotive},
and this type of squeezing, referred to as
ponderomotive squeezing~\cite{schnabel2017squeezed}, can be only used to evade back-action and thus it is less
extensive than externally injecting a squeezed light~\cite{corbitt2006squeezed} or generating intracavity squeezing via an OPA~\cite{wu1986generation}.
We note that in a recent experiment, the OPA effect was improved by 14 orders of magnitude via symmetry breaking
at a microcavity surface~\cite{Zhang2018OPA}.
Other approaches to generate intracavity squeezing include, e.g., Kerr
media~\cite{collett1984squeezing,bondurant1986reduction,yin2018enhanced} or
dissipative COM devices~\cite{xuereb2011dissipative,kronwald2014dissipative,bai2019qubit}.
We also note that by using mechanical squeezing~\cite{Squeezing2015Pirkkalainen,Noiseless2017Mika,Burd2019MPA,Qin2019npj},
the sensitivity of detecting small displacements
was improved by a factor of up to 7~\cite{Burd2019MPA}.

To realize weak-force sensing with OPA, we assume that only one
optical mode is coupled to the mechanical mode. Thus, the
Hamiltonian of the COM system can be described as~\cite{agarwal2016strong}:
\begin{align}
H_{0} & = \hbar\omega_{a}a^{\dagger}a+\frac{p^{2}}{2m}+\frac{1}{2}m\omega_{m}^2x^{2}+\hbar\frac{\omega_{a}}{L}xa^{\dagger}a \nonumber\\
& \quad\, + i\hbar G\left(e^{i\theta}a^{\dagger
2}e^{-2i\omega_{l}t}-e^{-i\theta}a^{2}e^{2i\omega_{l}t}\right),\label{H-interaction}
\end{align}
where $\omega_{a}$ denotes the optical resonant frequency;
$x$ and $p$ refer to the position and momentum operators of
the vibrating mechanical oscillator having an effective mass $m$
and an angular frequency $\omega_{m}$, respectively. In addition, $a$ and
$a^{\dagger}$ are the annihilation and creation operators of the
cavity mode, respectively. We have furthermore used $G$ to denote the
nonlinear gain of degenerate OPA with $\theta$ being the phase of
the pump field driving the OPA medium.
For the sake of simplicity, we define dimensionless position and
momentum operators of the mechanical mode as
$X=\left.x\middle/x_{\mathrm{zpf}}\right.$, $P=\left.p\middle/p_{\mathrm{zpf}}\right.$, where
$x_{\mathrm{zpf}}$ and $p_{\mathrm{zpf}}$ are the standard
deviations of the zero-point motion and momentum of the
oscillator, respectively,
$x_{\mathrm{zpf}}=\sqrt{\left.\hbar\middle/\left(m\omega_{m}\right)\right.}$,
$p_{\mathrm{zpf}}=\sqrt{\hbar m\omega_{m}}$, and the operators $X$
and $P$ satisfy the commutation relation $\left[X,P\right]=i$.
Thus, in a rotating frame, the Hamiltonian can be rewritten as follows:
\begin{align}
  H &= \hbar\Delta_{a}a^{\dagger}a+\frac{\hbar}{2}\omega_{m}\left(X^{2}+P^{2}\right)+\hbar g_{0}a^{\dagger}aX \nonumber\\
  & \quad\, +i\hbar G\left(e^{i\theta}a^{\dagger 2}-e^{-i\theta}a^{2}\right),\label{Hrotating-3}
\end{align}
where $g_{0}=\left.x_{\mathrm{zpf}}\omega_{a}\middle/L\right.$ represents the single
photon coupling strength of the COM interaction. The first and
second terms in Eq.~(\ref{Hrotating-3}) denote the sum of the free
energy of the cavity field and the mechanical mode without
external forces, respectively. And the last two terms respectively
describe the COM interaction and the contribution of an OPA.
By introducing dissipation and noise terms, the Heisenberg-Langevin equations can be written as:
\begin{align}
\dot{X}&=\omega_m P,\nonumber\\
\dot{P}&=-\omega_{m}X-\gamma_m P-g_0a^{\dagger}a+\sqrt{2\gamma_m}\left(f_{\mathrm{th}}+f_{\mathrm{ex}}\right),\nonumber\\
\dot{a}&=-\left(i\Delta_a+\frac{\kappa}{2}\right)a-ig_0X a+2G
e^{i\theta}a^{\dagger}+\sqrt{\kappa}a_{\mathrm{in}},
\label{mot}
\end{align}
where $\Delta_a=\omega_a-\omega_l$ is
the detuning of the input light frequency ($\omega_{a}$) with
respect to the cavity resonant frequency ($\omega_{l}$);
$a_{\mathrm{in}}$ characterizes the input field driving the
cavity, which fulfils~\cite{wimmer2014coherent,huang2018improving}: $\left\langle
a_{\mathrm{in}}\left(t\right)a^\dagger_{\mathrm{in}}\left(t'\right)\right\rangle
=\delta\left(t-t'\right)$. Moreover,
$f_{\mathrm{th}}=\left.\xi\middle/\sqrt{2\hbar m \gamma_m \omega_m}\right.$ and
$f_{\mathrm{ex}}=\left.F_{\mathrm{ex}}\middle/\sqrt{2\hbar m\gamma_m\omega_m}\right.$
are the scaled thermal and external forces with zero mean values,
respectively. The Brownian thermal noise operator $\xi$ obeys the
correlation
function~\cite{giovannetti2001phase}:
\begin{equation}
\left\langle\xi\left(t\right)\xi\left(t'\right)\right\rangle
=m\hbar\gamma_m\int\frac{\mathrm{d}\omega}{2\pi}e^{-i\omega\left(t-t'\right)}
\omega\left[\coth{\frac{\hbar\omega}{2k_BT}}+1\right],
\end{equation}
where $k_B$ is the Boltzmann constant and $T$ is the mirror
temperature of the thermal bath. Additionally, if the mechanical
quality factor is large, $Q=\left.\omega_m\middle/\gamma_m\right.\gg1$, the Brownian
noise $\xi\left(t\right)$ describes a Markovian process that is
delta-correlated~\cite{giovannetti2001phase}:
\begin{equation}\label{Brownian noise under Markovian approximation}
\tfrac
12\left\langle\xi\left(t\right)\xi\left(t'\right)+\xi\left(t'\right)\xi\left(t\right)\right\rangle\simeq
m\hbar\omega_m\gamma_m\left(2\bar{n}+1\right)\delta\left(t-t'\right),
\end{equation}
where
$\bar{n}=\left[\exp\left(\left.\hbar\omega_m\middle/k_BT\right.\right)-1\right]^{-1}$
is the mean thermal phonon number. Under the approximation of
thermal equilibrium and taking the classical limit
$\hbar\omega_m\ll k_BT$~\cite{bowen2015quantum}, the scaled
thermal force obeys $\left\langle
f_{\mathrm{th}}\left(t\right)f_{\mathrm{th}}\left(t'\right)
\right\rangle =\left(\left.k_B
T\middle/\hbar\omega_m\right.\right)\delta\left(t-t'\right)$~\cite{wimmer2014coherent}.

Linearization or even semiclassical approximation are
standard effective descriptions applied in case of strong optical
drives, which are valid in both theory~\cite{motazedifard2016force,huang2017robust} and experiments~\cite{safavi2013squeezed,Shomroni2018instability,basiri2019precision}.
Indeed, a strong quantum-optical drive can usually be
treated as a semiclassical parameter. Thus, an operator (a
q-number) describing such a drive can be replaced by a c-number.
For example, in standard description of homodyne detection, which
can be applied for quantum state tomography, a weak quantum signal
is driven (i.e., mixed on a beam splitter) by a strong-laser mode
(i.e., a local oscillator), which is described by a classical
parameter (see Appendix~\ref{Appendix D}).
Therefore, in case where the input optical field is a semiclassical coherent laser field~\cite{basiri2019precision}, i.e., in the strong driving regime $\left|\alpha\right|\gg 1$~\cite{huang2017robust} (see Appendix~\ref{Appendix B}), we can
expand each operator as the sum of its steady-state value and a
small fluctuation, i.e., $P=\bar{P}+\delta P$, $X=\bar{X}+\delta
X$, $a=\alpha+\delta a$, and
$a_{\mathrm{in}}=\alpha_{\mathrm{in}}+\delta a_{\mathrm{in}}$.
Here we choose the input field as the zero-phase reference, i.e.,
$\alpha_\mathrm{in}=\left|\alpha_\mathrm{in}\right|=\sqrt{\left.P_\mathrm{in}\middle/\hbar\omega_l\right.}$
with $P_{\mathrm{in}}$ being the input laser
power~\cite{bowen2015quantum}. By setting all the time derivatives
to zero, the steady-state values of the dynamical variables can be
obtained as $\bar{P}=0$, $\bar{X}=\left.-g_0 \alpha^*\alpha\middle/\omega_m\right.$
and
\begin{equation}
\alpha = \frac{\sqrt{\kappa}\alpha_\mathrm{in}}{2\sigma_{+}}\left(\kappa
-2i\Delta+4Ge^{i\theta}\right)=\left|\alpha\right|e^{i\phi},
\label{cs}
\end{equation}
where $\sigma_{\pm}=\left.\kappa^{2}\middle/4\right.\pm\Delta^{2}\mp 4G^{2}$ and $\Delta=\Delta_a+g_0 \bar{X}$ denotes the effective cavity
detuning. Therefore, the phase of the intracavity amplitude
becomes:
\begin{align}
\phi&=\arctan \left(
\frac{4G\sin\theta-2\Delta}{4G\cos\theta+\kappa}\right).
\label{phi}
\end{align}

We define the standard quadratures of the cavity field as $x_a=\left.\left( a+  a^{\dagger}\right)\middle/\sqrt{2}\right.$,
$p_a=\left.\left(a-a^{\dagger}\right)\middle/\left(\sqrt{2}i\right)\right.$.
These are the canonical (i.e., dimensionless) position ($x_a$) and
momentum ($p_a$) operators, also referred to as the amplitude and
phase quadratures, which are related to the electric and magnetic
fields of an optical mode, respectively.
For simplicity, we set the integral constants to zero and
neglect the higher-order terms $\delta a^\dagger\delta a$ and
$\delta X\delta a$, then the linearized equations can be written
as (a similar matrix form is given in
Ref.~\cite{agarwal2016strong}):
\begin{equation}
\dot{\mathrm{v}}=\mathrm{C}\mathrm{v}+\mathrm{A}\mathrm{v}_\mathrm{in}.
\label{mat}
\end{equation}
The operator vectors are $\mathrm{v}=
\begin{pmatrix}
X,&P,&x_a,& p_a
\end{pmatrix}
^\mathrm{T}$ and $\mathrm{v}_\mathrm{in}=
\begin{pmatrix}
0,&f_{\mathrm{in}},&x_a^{\mathrm{in}},&p_a^{\mathrm{in}}
\end{pmatrix}
^\mathrm{T}$,
where $f_{\mathrm{in}}=f_{\mathrm{th}}+ f_{\mathrm{ex}}$, and
superscript $\mathrm{T}$ represents the transpose of a matrix.
The quadratures of the input field ($x_{a}^{\mathrm{in}}$ and $p_{a}^{\mathrm{in}}$),
are defined analogously to $x_{a}$ and $p_{a}$.
The coefficient matrix $\mathrm{C}$ and the noise matrix $\mathrm{A}$ are given
in Appendix~\ref{Appendix B}.

In standard homodyne detection, which enables quantum state
tomography, a quantum optical field [in our case
$a_\mathrm{out}\left(t\right)$] is mixed with a strong classical
laser field $\alpha_{\mathrm{LO}}$ (i.e., a local oscillator, LO)
at a balanced beam splitter. The homodyne signal (say, the
photocurrent $i_{\mathrm{det}}$) corresponds to the difference of
the photon numbers at the two output ports of the beam splitter.
This signal is proportional to the generalized phase-dependent
quadrature operator $x_{a,\mathrm{out}}^{\varphi}$, as
follows~\cite{LeonhardtBook}:
\begin{align}
i_{\mathrm{det}} & = \left|\alpha_{\mathrm{LO}}\right|\left(
  e^{i\varphi}a^{\dagger}_{\mathrm{out}}+e^{-i\varphi}a_{\mathrm{out}}\right)\nonumber\\
& = \sqrt{2}\left|\alpha_{\mathrm{LO}}\right|x_{a,\mathrm{out}}^{\varphi},
\end{align}
where $\varphi$ and $\alpha_{\mathrm{LO}}$ respectively correspond to the phase and amplitude
of the local oscillator, which are arbitrary (see Appendix~\ref{Appendix D}
for more details).
Note that by changing the phase $\varphi$ of the local
oscillator, a complete quantum state tomography can be implemented
with this method~\cite{LeonhardtBook}.
As depicted in Figure~\ref{fig1}, homodyne
detection is a phase-referenced technique, where direct
measurement of an optical field produces a stochastic photocurrent
that is proportional to the rotated quadrature
$x^{\varphi}_\mathrm{a,out}$ and the measured quadrature is dependent
on the phase of the local oscillator $\alpha_\mathrm{LO}$.
We also note that the quadratures for arbitrary phases of the
local oscillator can be measured. Thus, homodyne tomography can be
performed to reconstruct the output field with
arbitrary phase~\cite{Phase2014Adam}. However, for brevity, we further study
only the output canonical momentum quadrature, i.e.,
$\varphi=\left.\pi\middle/2\right.$. Hence, the external force
can be expressed as:
\begin{equation}
F=\frac{1}{F_f\left(\omega\right)}x^{\left.\pi\middle/2\right.}_\mathrm{a,out}
=\frac{1}{F_f\left(\omega\right)}
\tilde{p}_a^\mathrm{out}= \tilde{f}_{\mathrm{ex}}+ \tilde{f}_{\mathrm{add}}.
\end{equation}
where $F_f\left(\omega\right)=gf_-\chi_{-}\chi_m \sqrt{2\kappa\gamma_m}$,
and $g=\sqrt{2}g_0\left|\alpha\right|$ represents the effective linearized
optomechanical coupling rate. Therefore, noise of the added force can be described by
\begin{equation}\label{induced force}
 f_{\mathrm{add}}\left(\omega\right)= \tilde{f}_{\mathrm{th}}+X_a \left(\omega\right)\tilde{x}_{a}^{\mathrm{in}}+P_a\left(\omega\right) \tilde{p}_{a}^{\mathrm{in}},
\end{equation}
where $\tilde{f}_{r}\equiv f_{r}\left(\omega\right)$, $\tilde{s}_{a}^{u}\equiv s_{a}^{u}\left(\omega\right)$, for $s=x,\,p$, $u=\mathrm{in,\,out}$, and $r=\mathrm{in,\,th,\,ex,\,add}$. Moreover,
the coefficients in Eq.~(\ref{induced force}) are given by:
\begin{equation}\label{key equations}
X_a\left(\omega\right)=\left.\left(\mu_{+}\lambda_{+}\chi_-\kappa\right)\middle/F_f\right.,\quad
P_a\left(\omega\right)=\left.\left(\chi_{-}\kappa-1\right)\middle/F_f\right. .
\end{equation}
The parameters $f_{\pm}$, $\chi_{\pm}$, $\mu_{\pm}$, and $\lambda_{\pm}$ are defined in Appendix~\ref{Appendix C}.

We use the symmetric part ($S_\mathrm{FF}$) of the added noise power spectral
density ($S_\mathrm{F}$) to characterize the sensitivity of the
force measurement, given by (see Appendix~\ref{Appendix D} for
more details):
\begin{align}
  S_{\mathrm{FF}}\left(\omega\right)
  &= \frac{S_{\mathrm{F}}\left(\omega\right)+S_{\mathrm{F}}\left(-\omega\right)}{2} \nonumber\\
  &=\underbrace{\frac{k_{B}T}{\hbar\omega_{m}}}_{\text{thermal noise}}+\underbrace{\frac{1}{2}\left|X_{a}\left(\omega\right)\right|^{2}}_{\text{backaction noise}}
  +\underbrace{\frac{1}{2}\left|P_{a}\left(\omega\right)\right|^{2}}_{\text{shot noise}},
  \label{symmetrised noise power spectral density}
\end{align}
where we have used the correlation functions and the bath cross-correlated terms of the measured symmetrized power spectral density are cancelled out. The first term of
$S_\mathrm{FF}\left(\omega\right)$ represents the thermal Brownian
noise. The second term is the back-action noise, which is
proportional to the input power $P_{\mathrm{in}}$ and the square
of the coupling strength $g^2$.
Very recently, Corbitt \emph{et al}. have presented a testbed for the
broadband measurement of quantum back action at room temperature~\cite{cripe2019measurement}.
The third term denotes the shot
noise that is inversely proportional to the input power
$P_{\mathrm{in}}$. Since beating the SQL in an optomechanical
sensor by cavity detuning has been studied in
Ref.~\cite{arcizet2006beating} and the highest parametric gain is
achieved at the cavity resonance~\cite{schnabel2017squeezed}, in
the following, we neglect thermal noise and other technical
noises and restrict our discussion to
the case of $\Delta=0$. We note that the idea of utilizing a
nondegenerate OPA-assisted COM to circumvent measured backaction
and surpass the SQL has been proposed in Ref.~\cite{wimmer2014coherent}.
This scheme is based on an antinoise process (using an oscillator
with an effective negative mass) via destructive quantum
interference, i.e., the so-called coherent quantum noise
cancellation (CQNC)~\cite{tsang2010coherent,tsang2012evading,gebremariam2019enhancing}.
Additionally, a quantum-mechanics-free subsystem (QMFS)~\cite{tsang2012evading}, proposed by Tsang and Caves~\cite{tsang2010coherent}, was first realized in Ref.~\cite{BAE2016Mika}.

%=============%
\begin{figure}[ht]
\centering
\includegraphics[width=3.2in]{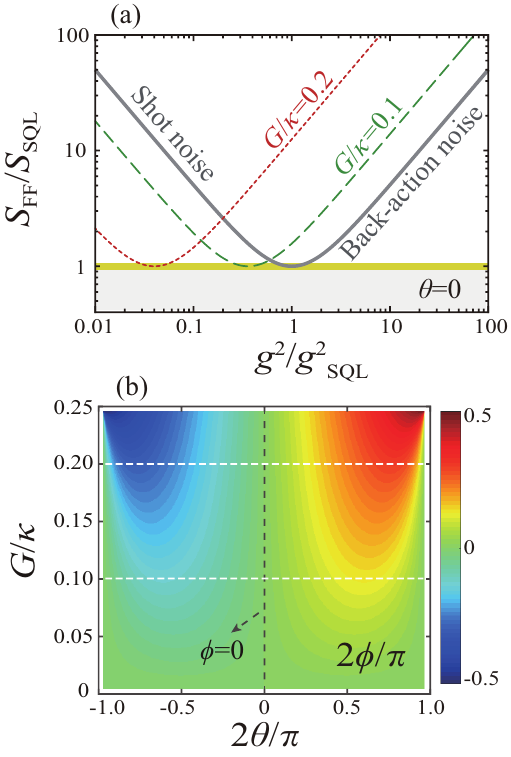}
\caption{(Color online) (a) Noise power spectral density
$\left.S_\mathrm{FF}\middle/S_{\mathrm{SQL}}\right.$ is plotted as a function of the
scaled square of the coupling strength $\left.g^2\middle/g^2_\mathrm{SQL}\right.$ and
of the OPA parametric gain $G$. The gray solid line corresponds to
the non-OPA case. (b) Phase $\phi$, given in Eq.~(\ref{phi}), of the
intracavity field versus the OPA parameters: the phase $\theta$
and gain $G$ in the units of the damping rate $\kappa$. Here we assume the resonance
condition $\Delta=0$.} \label{fig2}
\end{figure}
%--------------------------------------------------------------------------------------%

In order to compare force sensing in the presence and in the absence of OPA, we first
concern about force sensing of a standard COM system at a resonant
frequency. For such a scheme, the sensitivity cannot surpass the
SQL without an OPA [see Figure~\ref{fig2}(a)]. Moreover, if we
assume the linewidth of the cavity to be much larger than any
measurement frequency of interest, $\kappa\gg\omega$, the symmetrized noise
spectral density can be simplified as
follows~\cite{motazedifard2016force,wimmer2014coherent}:
\begin{equation}
S_\mathrm{FF}^\mathrm{st}=\frac{g^2}{\kappa\gamma_m}+\frac{1}{16}\frac{\kappa}
{g^2\gamma_m}\frac{1}{\left|\chi_m\right|^2},
\end{equation}
where the susceptibility of the mechanical oscillator ($\chi_{m}$)
has been defined in Appendix~\ref{Appendix B}. Thus, we obtain
$S_\mathrm{FF}^\mathrm{st}\geq \left(2\gamma_m\left|\chi_m\right|\right)^{-1}=S_{\mathrm{SQL}}$,
$g_\mathrm{SQL}^2=\left.\kappa\middle/\left(4\left|\chi_m\right|\right)\right.$.

%=============%
\begin{figure}[ht]
\centering
\includegraphics[width=3.2in]{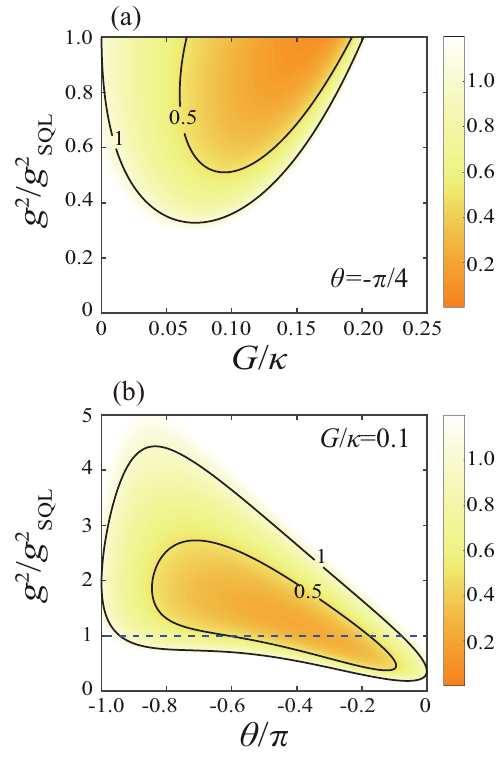}
\caption{(Color online) Noise power spectral density
$\left.S_\mathrm{FF}\middle/S_{\mathrm{SQL}}\right.$ as the functions of the scaled
square of the coupling strength, $\left.g^2\middle/g^2_\mathrm{SQL}\right.$, and the
parametric gain $G$ or the parametric phase $\theta$. The other
parameters are set as (a) $\theta=\left.-\pi\middle/4\right.$ and (b) $\left.G\middle/\kappa\right.=0.1$.}
\label{fig3}
\end{figure}
%=============%

Subsequently, we study the case of the zero pump phase
($\theta=0$) of the COM system with an OPA. In this case, the symmetrized
noise power spectral density $S_\mathrm{FF}$ can be reduced to:
\begin{equation}
S_\mathrm{FF}\left(\theta=0\right)=\frac{g^2\kappa}{4\gamma_m\left(\left.\kappa\middle/2\right.-2G\right)^2}
+\frac{\left(\left.\kappa\middle/2\right.-2G\right)^2}{4\kappa\gamma_mg^2\left|\chi_m\right|^2}.
\label{SF0}
\end{equation}
Therefore, we also have $S_\mathrm{FF}\left(\theta=0\right)\geq
\left.1\middle/\left(2\gamma_m\left|\chi_m\right|\right)\right.=S_{\mathrm{SQL}}$,
indicating that the detection sensitivity still cannot surpass the
SQL when $\theta=0$. As Figure~\ref{fig2}(a) shows, in the absence
of OPA ($\left.G\middle/\kappa\right.=0$), the minimum of
$\left.S_\mathrm{FF}\middle/S_{\mathrm{SQL}}\right.$ equals to 1 when the coupling
strength is $g=g_\mathrm{SQL}$. The optimal COM parameter
$g_\mathrm{opt}$ can be obtained by solving
$\left|X_a\right|=\left|P_a\right|$, which gives
$g_\mathrm{opt}^{\left(\theta=0\right)}=\left. \left|\kappa-4G\right|\middle/
\left(2\sqrt{\kappa\left|\chi_m\right|}\right) \right.$.
Furthermore, by the inspection of Eq.~(\ref{phi}) under the resonance condition
$\Delta=0$, it is easy to understand why the intracavity field
phases $\phi$ for $\pm\theta$, corresponding to symmetric
points on both sides of the dividing dashed line in Figure~\ref{fig2}(b),
have opposite signs. By comparing the phases $\phi$ for
$\theta=\pm\pi/2$ and $G/\kappa=1/4$, corresponding to the darkest
blue and darkest red points in Figure~\ref{fig2}(b), we find that the
intracavity field phases $\phi$ are $\pi/2$-shifted for the
opposite phases $\theta$ of the OPA pump for these special values
of $G=\kappa/4$ and $\Delta=0$. This conclusion can easily be
drawn by analyzing Eq.~(\ref{phi}) for the discussed parameters.

From the analyses made above, we have identified that it is
impossible for weak-force sensing to exceed the SQL with non-OPA
or $\theta=0$. As depicted in Figure~\ref{fig2}(a), owing to the
reduction of shot noise, $S_\mathrm{FF}$ first decreases with the
COM coupling strength increasing until the turning point
(corresponding to $S_\mathrm{FF}=S_{\mathrm{SQL}}$); then, the
backaction noise is dominant, leading $S_\mathrm{FF}$ to increase.
Hence, the SQL can be reached at the minimum point
$g=g_\mathrm{opt}$ and the optimum coupling strength can be
lowered by adjusting the OPA pump gain $G$. In calculations, we
use experimentally accessible parameters~\cite{bowen2015quantum},
i.e., $\left.\omega_l\middle/2\pi\right.=2\times 10^{14}\,\mathrm{Hz}$, $\left.\gamma_m\middle/2\pi\right.=1\,\mathrm{kHz}$,
$\left.\omega_m\middle/2\pi\right.=10\,\mathrm{MHz}$,
$\left.\kappa\middle/2\pi\right.=10\,\mathrm{MHz}$,
$\left.g_0\middle/2\pi\right.=100\,\mathrm{Hz}$ and $P_\mathrm{in}=700\,\mathrm{nW}$.

%=============%
\begin{figure*}[ht]
\centering
\includegraphics[width=6.8in]{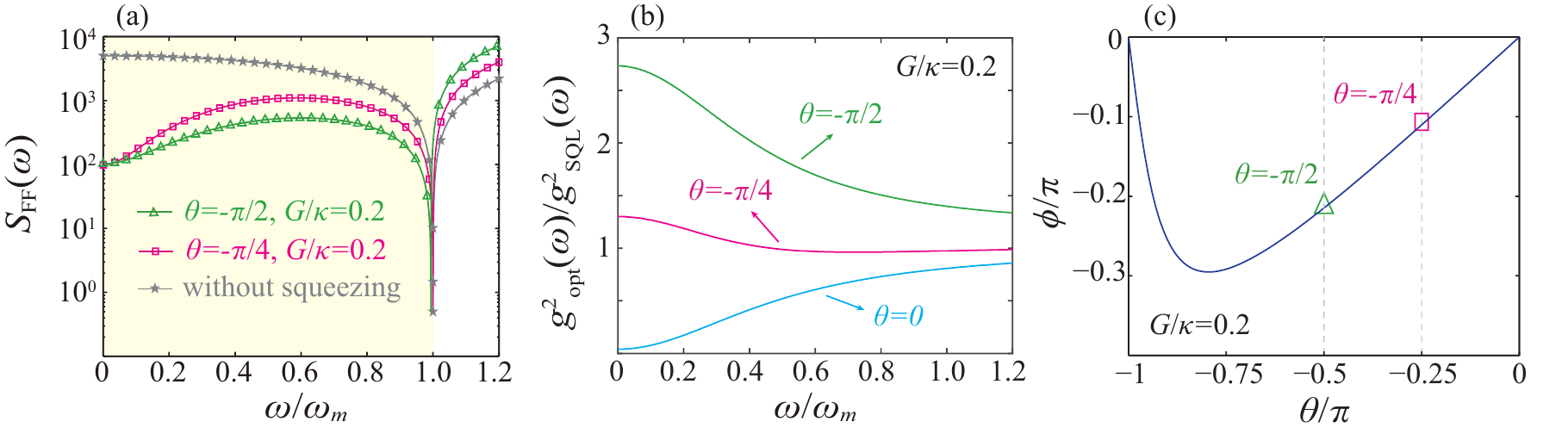}
\caption{(Color online) (a) Noise power spectral density $S_\mathrm{FF}(\omega)$
at the optimal power vs frequency $\omega$. The gray curve
corresponds to $S_\mathrm{SQL}(\omega)$ without squeezing. (b)
Optimal coupling strength as functions of the OPA pump phase and
the detection frequency. (c) Intracavity phase with varying the
OPA pump phase. The parametric gain is set as $\left.G\middle/\kappa\right.=0.2$.}
\label{fig4}
\end{figure*}
%The parametric phase is set as $\theta=0$. or detuning $\Delta$.
%=============%

To overcome the SQL, we study the impact of the OPA pump phase on
weak-force sensing. Figure~\ref{fig3} shows the sensitivity as the
functions of $\left.g^2\middle/g^2_\mathrm{SQL}\right.$ and the parametric gain $G$ in
the low-frequency domain. Specifically, as depicted in
Figure~\ref{fig3}(a), the sensitivity can be improved more than
twice with an OPA when choosing appropriate parameters and the
coupling strength required to increase the sensitivity becomes much
smaller than $g_\mathrm{SQL}$. Additionally, according to
Eqs.~(\ref{key equations}) and
(\ref{symmetrised noise power spectral density}), we are likely to enhance the
detection sensitivity by tuning $\sin\theta$ to satisfy
$2G\sin\theta+g^2\chi_m\cos^2\phi\approx 0$, for suppressing the
back-action noise to the limit. Furthermore, we utilize $\sin
\theta <0$ in Figure~\ref{fig3}(b) to reduce quantum noise and improve the measurement accuracy.
We note that very recently, optimal cavity squeezing up to $30\,\mathrm{dB}$ has been predicted theoretically~\cite{macri2019optimal}, and single label-free sensing of nanoparticles has been studied with a spinning resonator~\cite{Jing2018Nanoparticle} and exposed-core fiber~\cite{Mauranyapin2019Quantum}.
In particular, Miao \emph{et al}. have recently proposed a new fundamental limit to the precision of a gravitational-wave detector~\cite{Quantum2019Miao}, which provides an answer of how far we can push the detector sensitivity.

In Figure~\ref{fig4}, we consider the variation of the optimal symmetrized noise spectral density
$S_\mathrm{FF}(\omega)$ with the increase of the frequency
$\omega$, analogously similar to the corresponding illustration in
Ref.~\cite{wimmer2014coherent}. Here we set the OPA pump gain as
$\left.G\middle/\kappa\right.=0.2$.
As illustrated in Figure~\ref{fig4}(a), the SQL can be well
suppressed at frequencies below the mechanical resonance, i.e., we
can increase the force sensitivity by nearly two-orders of
magnitude in the low-frequency domain.
Note that recently, a quantum expander for gravitational-wave observatories has been demonstrated in Ref.~\cite{Korobko2019expander}, where quantum uncertainty can be squeezed even at high frequencies, while maintaining the low-frequency sensitivity unchanged, thus, expanding the detection bandwidth.
In Figures~\ref{fig4}(b)-\ref{fig4}(c), we depict the two key parameters, i.e., the optomechanical coupling strength $g_{\mathrm{opt}}$ and intracavity phase $\phi$, required for the minimum symmetrized noise spectral density, in the limits of different phases $\theta$ and frequencies $\omega$.

From what has been discussed above, it would be reasonable to
overcome the SQL by tuning the OPA pump phase. However, because of
this nonzero pump phase, the canonical position ($x_a$) and momentum ($p_a$) operators become
correlated, causing a loss of quantum efficiency which cannot be ignored
(see Appendix~\ref{Appendix B} for more details). For instance, the
dashed line in Figure~\ref{fig2}(b) illustrates the case of
$\theta=\phi=0$, in which $x_{a}$ and $p_{a}$ are decoupled,
circumventing quantum efficiency losses successfully, but
a loss of mechanical-mode information is inevitable in the regions
divided by the dashed line. Additionally, as the frequency increases, there is an
unavoidable quantum efficiency loss shown in Figure~\ref{fig4}. Nonetheless, in the
specific regime where $\kappa\gg\omega$ and $\Delta=0$, we are able
to achieve quantum noise reduction without losing mechanical-mode
information. In the limit of $\kappa\gg\omega$, according to the coefficients given in
Appendix~\ref{Appendix C}, we find:
\begin{equation}
  \mu_{\pm}\lambda_{\pm} \approx \frac{4G\sin\theta \pm 2g^{2}\chi_{m}\sin^{2}\left(\phi+\phi_{\pm}\right)}
  {\kappa \mp 4G\cos\theta \pm g^{2}\chi_{m}\sin\left(2\phi\right)},
\end{equation}
where $\phi_{\pm}=\left(1\pm 1\right)\left.\pi\middle/4\right.$. Hence, we obtain $\tan\phi=\mu_{-}\lambda_{-}$ and
$\cot\phi\neq\mu_{+}\lambda_{+}$, i.e., $f_{+}=0$ and $f_{-}\neq
0$. Utilizing the input-output relations, the
output canonical position quadrature $x_{a}^{\mathrm{out}}$ does not
contain $f_{\mathrm{in}}$; thus, the detected squeezed quadrature
carries all of the mechanical quantum information,
and the SQL can be reached or surpassed, as shown in Figures~\ref{fig2}-\ref{fig3},
whose parameters utilized in calculations obey the approximation $\kappa\gg\omega$.

\section{Conclusion}

In conclusion, we have investigated weak-force sensing in a
squeezed cavity and theoretically showed that (\rmnum{1})
the SQL cannot be surpassed in the case of $G=0$ or $\theta=0$,
(\rmnum{2}) the measurement precision of weak-force detection can
be remarkably improved at the coupling strength smaller than
$g_{\mathrm{SQL}}$ by tuning the parametric phase and gain, and
(\rmnum{3}) under the approximation of $\kappa\gg\omega$, quantum
noise can be reduced without losing mechanical-mode information. Our work provides new
insight in strengthening the sensitivity of a force sensor with
the assistance of intracavity squeezing, which can be also extended into other systems of quantum sensing with e.g., waveguide~\cite{Allen2019Passive,Mauranyapin2019Quantum}, interferometer~\cite{Ouyang2019On}, or parity-time ($\mathcal{PT}$) symmetric microcavity~\cite{Metrology2016Jing,ozdemir2019parity,Zhong2019EP}.
In the future, we
plan to extend our work to study the weak-force measurement with the help of two-mode squeezing or quantum entanglement~\cite{Palomaki710,takeuchi2019quantum,Stationary2019Barzanjeh},
squeezed mechanical modes~\cite{Suh1262,Burd2019MPA},
or squeezed sources in hybrid COM devices~\cite{li2019squeezed,Xiao2019squeezing}.

\section*{Acknowledgments}
This work was supported by the National Natural Science Foundation of
China (NSFC) (Grants Nos.~11474087, and 11774086), the Key Program of
NSFC (Grant No.~11935006), and the HuNU Program for Talented Youth.

\begin{appendix}
\appendix
\renewcommand{\appendixname}{APPENDIX}
\section{DERIVATION OF THE EFFECTIVE HAMILTONIAN}\label{Appendix A}
\makeatletter
\renewcommand\theequation{A\@arabic\c@equation }
\makeatother \setcounter{equation}{0}

In our system, the cavity with
resonant frequency $\omega_a$ and damping rate $\kappa$, is driven
by an input beam $a_\mathrm{in}$ at frequency $\omega_l$. The left
mirror is movable, which supports a mechanical mode with frequency
$\omega_m$ and damping rate $\gamma_m$.
An external force
$F_{\mathrm{ex}}$ is applied on the left-hand mirror and the
cavity adjoins the movable mirror with coupling strength
$g_0=\left.x_{\mathrm{zpf}}\omega_a\middle/L\right.$, where $L$ is the length of the
cavity. When a pump field at frequency $2\omega_l$ interacts with
a second-order nonlinear optical crystal, the output frequency
becomes $\omega_l$~\cite{huang2017robust}. The nonlinear gain $G$
of the degenerate OPA with pump phase $\theta$ is proportional to
the pump field.
By transforming into a frame rotating at the incoming laser frequency $\omega_{l}$
with $\Delta_{a}=\omega_{a}-\omega_{l}$, Hamiltonian
(\ref{H-interaction}) can be rewritten as:
\begin{equation}
H = -i\hbar
U\frac{\mathrm{d}U^{\dagger}}{\mathrm{d}t}+UH_{0}U^{\dagger},
\label{Hrotating-1}
\end{equation}
via the unitary transformation
$U\left(t\right)=\exp\left(i\omega_{l}a^{\dagger}at\right)$. Using
the relation
\begin{equation}
  e^{A}Be^{-A}=B+\left[A,B\right]+\frac{1}{2!}\left[A,\left[A,B\right]\right]+\cdots, \label{the Hadamard lemma}
\end{equation}
we obtain~\cite{huang2009enhancement,huang2009normal}:
\begin{align}
  H &= \hbar\Delta_{a}a^{\dagger}a+\hbar\frac{\omega_{a}}{L}xa^{\dagger}a+\frac{p^{2}}{2m}+\frac{1}{2}m\omega_{m}^2x^{2} \nonumber\\
  & \quad\, +i\hbar G\left(e^{i\theta}a^{\dagger 2}-e^{-i\theta}a^{2}\right), \label{Hrotating-2}
\end{align}
where $m$ stands for the effective
mass of the mechanical mode, while $x$ and $p$ represent the position
and momentum operators, respectively.

In the Heisenberg picture, the dynamics of an operator
$\mathcal{O}$ of a quantum system can be determined via the
Heisenberg equation of motion:
\begin{equation}
\dot{\mathcal{O}}\left(t\right)=\frac{1}{i\hbar}\left[\mathcal{O}\left(t\right),H\left(t\right)\right].
\end{equation}
By introducing dissipation and noise terms, we have:
\begin{align}
  \dot{X} &= \omega_{m}P, \nonumber\\
  \dot{P} &= -\omega_{m}X-\gamma_{m}P-g_{0}a^{\dagger}a
   +\sqrt{2\gamma_{m}} \left( f_{\mathrm{th}}+f_{\mathrm{ex}} \right), \nonumber\\
  \dot{a} &= -i\left( \Delta_{a}+g_{0}X \right)a +2Ge^{i\theta}a^{\dagger}
   -\frac{\kappa}{2}a+\sqrt{\kappa}a_{\mathrm{in}},\label{motionEq}
\end{align}
where $\kappa$ and $\gamma_{m}$ are the decay rates of the optical
cavity and the mechanical oscillator, respectively.
$X=\left.x\middle/x_{\mathrm{zpf}}\right.$ and $P=\left.p\middle/p_{\mathrm{zpf}}\right.$ denote the dimensionless displacement and momentum operators of the mechanical mode, where
$x_{\mathrm{zpf}}=\sqrt{\left.\hbar\middle/\left(m\omega_m\right)\right.}$ and
$p_{\mathrm{zpf}}=\sqrt{\hbar m\omega_m}$ are the zero-point
position and momentum fluctuations, respectively.
The input noise operator $a_{\mathrm{in}}\left(t\right)$ fulfills the
correlation relations~\cite{wimmer2014coherent,huang2018improving}:
  \begin{align}\label{inputRelation}
  \left\langle a_{\mathrm{in}}^{\dagger}\left(t\right)a_{\mathrm{in}}\left(t'\right)\right\rangle
  &= 0, \nonumber\\
  \left\langle a_{\mathrm{in}}\left(t\right)a_{\mathrm{in}}^{\dagger}\left(t'\right) \right\rangle
  &= \delta\left(t-t'\right), \nonumber\\
  \left\langle a_{\mathrm{in}}\left(t\right)a_{\mathrm{in}}\left(t'\right)\right\rangle
  &= \left\langle a_{\mathrm{in}}^{\dagger}\left(t\right)a_{\mathrm{in}}^{\dagger }\left(t'\right)\right\rangle=0.
\end{align}
Moreover, $f_{\mathrm{th}}$ and $f_{\mathrm{ex}}$ represent the
scaled thermal and external forces given, respectively, by:
\begin{equation}
  f_{\mathrm{th}}=\frac{\xi}{\sqrt{2\hbar m\gamma_{m}\omega_{m}}},\quad
  f_{\mathrm{ex}}=\frac{F_{\mathrm{ex}}}{\sqrt{2\hbar m\gamma_{m}\omega_{m}}},\label{scaled forces}
\end{equation}
where $\xi$ and $F_{\mathrm{ex}}$ are the corresponding thermal
and external forces, respectively.

When the mechanical resonator is in thermal equilibrium at
environment temperature $T$, the Bose-Einstein statistics
determines the occupancy probability $p\left(n\right)$ of each
energy level, given by:
\begin{equation}
p\left(n\right)=\delta^{n}\left(1-\delta\right),
\end{equation}
where
$\delta=\exp\left[\left.-\hbar\omega_{m}\middle/\left(k_{B}T\right)\right.\right]$.
Therefore, the mean number $\bar{n}$ of phonons in thermal
equilibrium is
\begin{equation}
  \bar{n} = \sum_{n=0}^{\infty} np\left(n\right)=\left(\delta^{-1}-1\right)^{-1}.\label{mean number of phonons}
\end{equation}
In the high temperature limit of $k_{B}T\gg\hbar\omega_{m}$,
Eq.~(\ref{mean number of phonons}) can be simplified to the
familiar expression:
\begin{equation}
  \bar{n}\approx\frac{k_{B}T}{\hbar\omega_{m}},\label{classical mean number of phonons}
\end{equation}
and according to Eq.~(\ref{Brownian noise under Markovian
approximation}), the thermal noise force satisfies the correlation
function as follows~\cite{wimmer2014coherent}:
\begin{equation}
  \left\langle \xi\left(t\right)\xi\left(t'\right) \right\rangle
  =2m\gamma_{m}k_{B}T\delta\left(t-t'\right). \label{thermal force correlation function}
\end{equation}
 Combining Eq.~(\ref{scaled forces}) with Eq.~(\ref{thermal force correlation function}), we obtain:
\begin{equation}
  \left\langle f_{\mathrm{th}}\left(t\right)f_{\mathrm{th}}\left(t'\right) \right\rangle
     = \frac{k_{B}T}{\hbar\omega_{m}}\delta\left(t-t'\right).\label{scaled thermal force correlation function}
\end{equation}
Hence, Eq.~(\ref{scaled thermal force correlation function})
becomes~\cite{wimmer2014coherent}:
\begin{equation}
  \left\langle f_{\mathrm{th}}\left(t\right)f_{\mathrm{th}}\left(t'\right) \right\rangle
  =\bar{n}\delta\left(t-t'\right). \label{f-th-correlation}
\end{equation}

\section{LINEAR RESPONSE}\label{Appendix B}
\makeatletter
\renewcommand\theequation{B\@arabic\c@equation }
\makeatother \setcounter{equation}{0}

To solve the nonlinear Heisenberg-Langevin equations in Eq.~(\ref{mot}), we linearize the operators around the steady-state values
$\left(\bar{X},\,\bar{P},\,\alpha,\,\bar{f}_{\mathrm{th}},\,\bar{f}_{\mathrm{ex}},\,
\alpha_{\mathrm{in}}\right)$, i.e., insert the ansatz
  \begin{align}
  X &= \bar{X}+\delta X, & P &= \bar{P}+\delta P, \nonumber\\
  a &= \alpha+\delta a, & f_{\mathrm{th}} &= \bar{f}_{\mathrm{th}}+\delta f_{\mathrm{th}}, \nonumber\\
  f_{\mathrm{ex}} &= \bar{f}_{\mathrm{ex}}+\delta f_{\mathrm{ex}}, & a_{\mathrm{in}} &= \alpha_{\mathrm{in}}+\delta a_{\mathrm{in}},
\end{align}
into Eq.~(\ref{motionEq}), and retain only the first-order terms, then we obtain:
  \begin{align}\label{H-L eq}
 \dot{X} &= \omega_{m} P, \nonumber\\
 \dot{P} &= -\omega_{m} X-\gamma_{m} P-g_{0}\left(\alpha^{*} a+\alpha a^{\dagger}\right)  +\sqrt{2\gamma_{m}}f_{\mathrm{in}},\nonumber\\
 \dot{a} &= -\left(i\Delta_{a}+\frac{\kappa}{2}\right)a-ig_{0}\left(\bar{X} a+\alpha X\right)+2Ge^{i\theta} a^{\dagger}+\sqrt{\kappa} a_{\mathrm{in}}.
\end{align}
where $f_{\mathrm{in}}=f_{\mathrm{th}}+f_{\mathrm{ex}}$ and we
consider the thermal and external noises average to
$0$. For simplicity, we set the
integral constants to zero.

When all time derivatives vanish, the steady-state values fulfill
the self consistent equations:
\begin{align}\label{steady-state eq}
 0 &= \omega_{m}\bar{P}, \nonumber\\
 0 &= -\omega_{m}\bar{X}-\gamma_{m}\bar{P}-g_{0}\alpha^{*}\alpha, \nonumber\\
 0 &= -\left(i\Delta_{a}+\frac{\kappa}{2}\right)\alpha-ig_{0}\bar{X}\alpha+2Ge^{i\theta}\alpha^{*}
 +\sqrt{\kappa}\alpha_{\mathrm{in}}.
\end{align}
The steady-state solution of Eq.~(\ref{steady-state eq}) can be given by:
$\bar{P}=0$, $\bar{X} = \left.-g_{0}n_{a}\middle/\omega_{m}\right.$, and
\begin{equation}
\alpha = \frac{\sqrt{\kappa}}{2\sigma_{+}}
\left[\left(\kappa-2i\Delta\right)\alpha_{\mathrm{in}}
+4Ge^{i\theta}\alpha_{\mathrm{in}}^{*}\right] =
\left|\alpha\right|e^{i\phi}, \label{static solution1}
\end{equation}
where $\sigma_{\pm}=\left.\kappa^{2}\middle/4\right.\pm\Delta^{2}\mp 4G^{2}$ and $\phi$ is the
phase of the intracavity field. We have defined the effective
detuning ($\Delta = \Delta_{a}+g_{0}\bar{X}$) and the mean
intracavity photon number ($n_{a}=\left|\alpha\right|^{2}$).

Since phase is relative while phase difference is absolute, we
focus on the phase difference between the external and internal
optical fields and choose the incoming field as the zero phase
reference, i.e.,
$\alpha_{\mathrm{in}}=\left|\alpha_{\mathrm{in}}\right|=\sqrt{\left.P_{\mathrm{in}}\middle/\hbar\omega_{l}\right.}$,
where $P_{\mathrm{in}}$ denotes the input laser
power~\cite{bowen2015quantum}. Thus, Eq.~(\ref{static solution1})
can be simplified to the following expression:
\begin{equation}
  \alpha = \frac{\sqrt{\kappa}\alpha_{\mathrm{in}}}{2\sigma_{+}}
  \left(\kappa-2i\Delta+4Ge^{i\theta}\right)
    = \left|\alpha\right|e^{i\phi}. \label{static solution2}
\end{equation}
The assumption that $\alpha_{\mathrm{in}}$ is real, makes the phase
$\phi$ dependent on $\kappa$, $ G$, $\Delta$, and $\theta$, as
follows:
\begin{equation}
  \phi=\arctan\left(\frac{4G\sin\theta-2\Delta}{4G\cos\theta+\kappa}\right). \label{definition of phi}
\end{equation}
Furthermore, $\left|\alpha\right|$ can be solved from Eq.~(\ref{static solution2}):
\begin{equation}
  \left|\alpha\right|=\left|\frac{\sqrt{\kappa}\alpha_{\mathrm{in}}}{\sigma_{+}}\right|
  \sqrt{\sigma+2G\left(\kappa\cos\theta-2\Delta\sin\theta\right)},
\end{equation}
where $\sigma=\left.\kappa^{2}\middle/4\right.+\Delta^{2}+4G^{2}$.
In the limit of $\Delta=0$, we use experimentally accessible parameters, which have been listed
in Section~\ref{Section 2}, to estimate the magnitude of $\left|\alpha\right|$:
\begin{equation}
  \left|\alpha\right|\geq \left|\frac{2\sqrt{\kappa}\alpha_{\mathrm{in}}}{\kappa+4G}\right|
  \sim 10^{2}\gg 1.
\end{equation}
Hence, our calculations satisfy the strong-driving condition $\left|\alpha\right|\gg 1$~\cite{huang2017robust} and this linearized model has been proved valid and physically reasonable for our COM system.

To obtain the solutions of Eq.~(\ref{H-L eq}), we define the
quadratures of input/output fields as $x_{a}^{u} =
\left.\left(a_{u}+a^{\dagger}_{u}\right)\middle/\sqrt{2}\right.$ and $p_{a}^{u} =
\left.\left(a_{u}- a^{\dagger}_{u}\right)\middle/\left(\sqrt{2}i\right)\right.$,
where $u=\mathrm{in,\,out}$,
analogously similar to the canonical position and momentum operators (given in Section~\ref{Section 2}).
Then, we have the linearized Heisenberg-Langevin equations given in Eq.~(\ref{mat}),
with the coefficients:
\begin{align}\label{coefficient and noise matrix}
\mathrm{A} &=
  \begin{pmatrix}
    0 & 0 & 0 & 0 \\
    0 & \sqrt{2\gamma_m} & 0 & 0 \\
    0 & 0 & \sqrt{\kappa} & 0 \\
    0 & 0 & 0 & \sqrt{\kappa} \\
  \end{pmatrix},\nonumber\\
\mathrm{C} &=
  \begin{pmatrix}
    0 & \omega_m & 0 & 0\\
    -\omega_m & -\gamma_m & -g\cos\phi & -g\sin\phi\\
    g\sin\phi & 0 & C_{-} & S_{+}\\
    -g\cos\phi & 0 & S_{-} & -C_{+}\\
  \end{pmatrix},
\end{align}
where $C_{\pm}=2G\cos\theta\pm\left.\kappa\middle/2\right.$,
$S_{\pm}=2G\sin\theta\pm\Delta$, and
$g=\sqrt{2}g_0\left|\alpha\right|$ is the effective linearized
optomechanical coupling strength.
Using the Fourier transform
of Eq.~(\ref{mat}), we obtain the linearized Heisenberg-Langevin equations in the frequency domain:
\begin{equation}\label{omat}
-i\omega\tilde{\mathrm{v}}=\mathrm{C}\tilde{\mathrm{v}}+\mathrm{A}\tilde{\mathrm{v}}_\mathrm{in},
\end{equation}
where $\tilde{\mathrm{v}}\equiv\mathrm{v}\left(\omega\right)=
\begin{pmatrix}
\tilde{X},&\tilde{P},&\tilde{x}_a,&\tilde{p}_a
\end{pmatrix}
^\mathrm{T}$, $\tilde{\mathrm{v}}_\mathrm{in}\equiv\mathrm{v}_\mathrm{in}\left(\omega\right)=
\begin{pmatrix}
0,&\tilde{f}_{\mathrm{in}},&\tilde{x}_a^{\mathrm{in}},&\tilde{p}_a^{\mathrm{in}}
\end{pmatrix}
^\mathrm{T}$, and
$\tilde{f}_{\mathrm{in}}\equiv\tilde{f}_{\mathrm{in}}\left(\omega\right)$;
$\tilde{o}\equiv o\left(\omega\right)$, $\tilde{s}_{a}\equiv
s_{a}\left(\omega\right)$, and $\tilde{s}_{a}^{\mathrm{in}}\equiv
s_{a}^{\mathrm{in}}\left(\omega\right)$, for $o=X,\,P$, and $s=x,\,p$.
From Eqs.~(\ref{mat}) and (\ref{coefficient and noise matrix}),
we can see when inserting a degenerate OPA medium into the
Fabry-P\'{e}rot cavity, the canonical position ($x_a$) and momentum ($p_a$) operators are decoupled only if $S_{\pm}=0$.
Specifically, in the dashed line showed in Fig.~\ref{fig2}(b), substituting $\Delta=\theta=0$ ($S_{\pm}=0$) into Eq.~(\ref{omat}), we find that all the information is imprinted on the squeezed canonical momentum quadrature:
$\tilde{x}_{a} = \rho_{-}\sqrt{\kappa}\tilde{x}_{a}^{\mathrm{in}}$, and
\begin{align}\label{dashed line}
  \tilde{p}_{a} & = g^{2}\chi_{m}\rho_{+}\rho_{-}\sqrt{\kappa}\tilde{x}_{a}^{\mathrm{in}}
  +\rho_{+}\sqrt{\kappa}p_{a}^{\mathrm{in}} \nonumber\\
  & \quad\, -g\chi_{m}\rho_{+}\sqrt{2\gamma_{m}}\tilde{f}_{\mathrm{in}},
\end{align}
where $\rho_{\pm}=\left(\chi^{-1}\pm 2G\right)^{-1}$,
and the susceptibilities of the cavity field and the mechanical
oscillator are respectively defined
as~\cite{huang2018improving}:
\begin{align}\label{susceptibility}
\chi\left(\omega\right)&=\left(\left.\kappa\middle/2\right.-i\omega\right)^{-1},\nonumber\\
\chi_m\left(\omega\right)&=\omega_m
\left(\omega_m^2-\omega^2-i\omega\gamma_m\right)^{-1}.
\end{align}
Clearly, there are no correlations between this squeezed quadrature
and its canonically conjugated quadrature $\tilde{x}_{a}$.
However, on the both sides of the dividing line in Fig.~\ref{fig2}(b), loss of mechanical-mode information inevitably exists, for the two quadratures are correlated.
Therefore, we can conclude that under the resonance condition $\Delta=0$, quantum efficiency losses are dependent on the coupling of the quadratures $x_a$ and $p_a$ (i.e., the coefficients $S_{\pm}$ or the phase $\theta$ of the OPA pump).
Our analysis of the stability conditions for the matrix $\mathrm{C}$, in Eq.~(\ref{coefficient and noise matrix}), and the input-output relations are given in Appendix~\ref{Appendix C}.

\section{STABILITY CONDITIONS AND INPUT-OUTPUT RELATIONS}\label{Appendix C}
\makeatletter
\renewcommand\theequation{C\@arabic\c@equation }
\makeatother \setcounter{equation}{0}

The system is stable only if all
the eigenvalues $\lambda$ of the matrix $\mathrm{C}$ have negative
real parts~\cite{jeffrey2007table}. It is well known that the
characteristic equation $\left|\mathrm{C}-\lambda
\mathrm{I}\right|=0 $ can be reduced to:
\begin{equation}
  \lambda^4+\mathrm{C}_3\lambda^3+\mathrm{C}_2\lambda^2+\mathrm{C}_1\lambda+\mathrm{C}_0=0.
\end{equation}
Hence, we obtain the stability conditions of the system from
the Routh-Hurwitz criterion~\cite{jeffrey2007table}:
\begin{align}
0&<\mathrm{C}_3,\nonumber\\
0&<\mathrm{C}_3\mathrm{C}_2-\mathrm{C}_1,\nonumber\\
0&<\mathrm{C}_3\mathrm{C}_2\mathrm{C}_1-\left(\mathrm{C}_1^2+\mathrm{C}_3^2\mathrm{C}_0\right).
\label{RH}
\end{align}
Specifically, all the external parameters should be chosen to
satisfy the stability conditions in Eq.~(\ref{RH}), where the
coefficients of the characteristic equation can be given by:
\begin{align}
  \mathrm{C}_3 &= \kappa+\gamma_{m}, \nonumber\\
  \mathrm{C}_2 & = \omega_{m}^{2}+\kappa\gamma_{m}+\sigma_{+}, \nonumber\\
  \mathrm{C}_1 & = \kappa\omega_{m}^{2}+\sigma_{+}\gamma_{m}, \nonumber\\
  \mathrm{C}_0 & = \sigma_{+}\omega_{m}^{2}+2g^{2}G\omega_{m}\sin\left(2\phi-\theta\right)
  -g^{2}\omega_{m}\Delta.
\end{align}

By solving Eq.~(\ref{omat}), we obtain the quadratures $\tilde{x}_a$ and $\tilde{p}_a$:
  \begin{align}\label{important coefficients}
  \tilde{x}_{a} &= gf_{+}\chi_{+}\chi_{m}\sqrt{2\gamma_{m}}\tilde{f}_{\mathrm{in}}
  +\chi_{+}\sqrt{\kappa}\tilde{x}_{a}^{\mathrm{in}} \nonumber\\
  & \quad\, +\mu_{-}\lambda_{-}\chi_{+}\sqrt{\kappa}\tilde{p}_{a}^{\mathrm{in}}, \nonumber\\
  \tilde{p}_{a} &= gf_{-}\chi_{-}\chi_{m}\sqrt{2\gamma_{m}}\tilde{f}_{\mathrm{in}}
  +\chi_{-}\sqrt{\kappa}\tilde{p}_{a}^{\mathrm{in}} \nonumber\\
  & \quad \,+\mu_{+}\lambda_{+}\chi_{-}\sqrt{\kappa}\tilde{x}_{a}^{\mathrm{in}},
\end{align}
where $\tilde{f}_{r}\equiv f_{r}\left(\omega\right)$, $\tilde{s}_{a}^{u}\equiv s_{a}^{u}\left(\omega\right)$, for $s=x,\,p$, $u=\mathrm{in,\,out}$, and $r=\mathrm{in,\,th,\,ex,\,add}$. Moreover, the coefficients in Eq.~(\ref{important coefficients}) are given by:
\begin{align}\label{important variables}
f_-\left(\omega\right)&=\mu_{+}\lambda_{+}\sin\phi-\cos\phi,\nonumber\\
f_+\left(\omega\right)&=\sin\phi-\mu_{-}\lambda_{-}\cos\phi,\nonumber\\
\chi_{\pm}\left(\omega\right)&=\left(\lambda_{\pm}^{-1}-\mu_{+}\mu_{-}\lambda_{\mp}\right)^{-1},\nonumber\\
\mu_{\pm}\left(\omega\right)&=\mp\Delta+2G\sin\theta\pm g^2\chi_m\cos^2\phi,\nonumber\\
\lambda_{\pm}\left(\omega\right)&=\left[\chi^{-1}\mp 2G\cos \theta\pm
\tfrac 12 g^2\chi_m\sin\left(2\phi\right)\right]^{-1}.
\end{align}
Using the input-output relations
$\tilde{x}_{a}^{\mathrm{out}}=\sqrt{\kappa}\tilde{x}_{a}-\tilde{x}_{a}^{\mathrm{in}}$,
$\tilde{p}_{a}^{\mathrm{out}}=\sqrt{\kappa}\tilde{p}_{a}-\tilde{p}_{a}^{\mathrm{in}}$,
the quadratures of the output fields are given by:
\begin{align}\label{x-a-out,p-a-out}
  \tilde{x}_{a}^{\mathrm{out}} &= gf_{+}\chi_{+}\chi_{m}\sqrt{2\kappa\gamma_{m}}
  \tilde{f}_{\mathrm{in}} +\mu_{-}\lambda_{-}\chi_{+}\kappa \tilde{p}_{a}^{\mathrm{in}}\nonumber\\
  & \quad\, +\left(\chi_{+}\kappa-1\right)\tilde{x}_{a}^{\mathrm{in}}, \nonumber\\
  \tilde{p}_{a}^{\mathrm{out}} &= gf_{-}\chi_{-}\chi_{m}\sqrt{2\kappa\gamma_{m}}
  \tilde{f}_{\mathrm{in}} +\mu_{+}\lambda_{+}\chi_{-}\kappa \tilde{x}_{a}^{\mathrm{in}}\nonumber\\
  & \quad\, +\left(\chi_{-}\kappa-1\right)\tilde{p}_{a}^{\mathrm{in}}.
\end{align}

\section{MOTIONAL NOISE SPECTRUM}\label{Appendix D}
\makeatletter
\renewcommand\theequation{D\@arabic\c@equation }
\makeatother \setcounter{equation}{0}

Herein, we describe the outcoming optical field and the laser field with the
annihilation operators $a_{\mathrm{out}}\left(t\right)$ and $a_{\mathrm{LO}}\left(t\right)$. When they interact with a 50/50 beam splitter simultaneously,
extracavity photons can be transformed into photoelectrons,
generating two photocurrents:
\begin{equation}
  i_{k}\left(t\right) = n_{k}\left(t\right)
  =a'^{\dagger}_{k}\left(t\right)a'_{k}\left(t\right),
\end{equation}
where $k=1,\,2$ and $n_{k}$ are the photon number operators measured in the two detectors. Making the parametric approximation for the photon number of the laser field, $\left\langle n_{\mathrm{LO}}\right\rangle\gg 1$, we have $a_{\mathrm{LO}}\approx\alpha_{\mathrm{LO}}$, where $\alpha_{\mathrm{LO}}$ denotes the amplitude of the local oscillator. In homodyne detection, the relations between $a_{\mathrm{out}}$, $a_{\mathrm{LO}}$ and $a'_{k}$ are expressed as:
\begin{equation}
  a'_{k} = \frac{1}{\sqrt{2}}\left[a_{\mathrm{out}}+\left(-1\right)^{k}a_{\mathrm{LO}}\right].
\end{equation}
We measure the difference of the intensities, which can be written as:
\begin{align}
  i_{\mathrm{det}} & = i_{2}-i_{1} = \alpha^{*}_{\mathrm{LO}}a_{\mathrm{out}}
  +\alpha_{\mathrm{LO}}a^{\dagger}_{\mathrm{out}}\nonumber\\
  & = \left|\alpha_{\mathrm{LO}}\right|\left(e^{-i\varphi}a_{\mathrm{out}}
  +e^{i\varphi}a^{\dagger}_{\mathrm{out}}\right),
\end{align}
where we have used $a_{\mathrm{LO}}\approx\alpha_{\mathrm{LO}}$, and $\varphi$ represents the phase of the local oscillator. Thus, the values of $\varphi$ and $\left|\alpha_{\mathrm{LO}}\right|$ can be arbitrary.

We define dimensionless quadrature operators $x_{a,\mathrm{out}}^{\varphi}$ and
$p_{a,\mathrm{out}}^{\varphi}$ rotated by a phase angle $\varphi$ from
$x_{a}^{\mathrm{out}}$ and $p_{a}^{\mathrm{out}}$ as follows:
\begin{equation}
\begin{pmatrix}
  x_{a,\mathrm{out}}^{\varphi} \\
  p_{a,\mathrm{out}}^{\varphi}
\end{pmatrix}
= \begin{pmatrix}
  \cos\varphi & \sin\varphi \\
  -\sin\varphi & \cos\varphi
\end{pmatrix}
\begin{pmatrix}
  x_{a}^{\mathrm{out}} \\
  p_{a}^{\mathrm{out}}
\end{pmatrix},
\end{equation}
which obey the commutation relation $\left[x_{a,\mathrm{out}}^{\varphi},p_{a,\mathrm{out}}^{\varphi}\right]=i$.
Thus, we obtain the detected field operator:
\begin{equation}
  i_{\mathrm{det}}\left(t\right)=\sqrt{2}\left|\alpha_{\mathrm{LO}}\right|x_{a,\mathrm{out}}^{\varphi}.
\end{equation}
We use $\psi$ to describe the phase of the outcoming field, and Eq.~(\ref{static solution2}) yields for the expression of $\psi$ via the input-output relation:
\begin{align}
  \alpha_{\mathrm{out}} &= \left|\alpha_{\mathrm{out}}\right|e^{i\psi}
  =\sqrt{\kappa}\alpha-\alpha_{\mathrm{in}} \nonumber\\
  &= \frac{\kappa\alpha_{\mathrm{in}}}{2\sigma_{+}}
  \left(\kappa-2i\Delta+4Ge^{i\theta}\right)
  -\alpha_{\mathrm{in}}. \label{static solution3}
\end{align}
For the assumption that $\alpha_{\mathrm{in}}$ is real, $\psi$ is
dependent on $\kappa$, $G$, $\Delta$, and $\theta$,
\begin{equation}\label{definition of psi}
  \psi=\arctan\left(\frac{2G\kappa\sin\theta-\Delta\kappa}
  {2G\kappa\cos\theta+\sigma_{-}}\right),
\end{equation}
where $\sigma_{-}$ has been defined below Eq.~(\ref{static solution1}).

In the specific case of an optomechanical system without detuning
and OPA ($\Delta=G=0$), we obtain $\phi=\psi=0$ from
Eqs.~(\ref{definition of phi}) and (\ref{definition of psi}), therefore, $\mu_{\pm}$, $\lambda_{\pm}$, and $\chi_{\pm}$ can be simplified to:
$\mu_{+} = g^{2}\chi_{m}$,
$\mu_{-} = 0$, and $\lambda_{\pm}= \chi_{\pm}=\chi=\left(\left.\kappa\middle/2\right.-i\omega\right)^{-1}$.
Thus, Eq.~(\ref{x-a-out,p-a-out}) can be written as
$\tilde{x}_{a}^{\mathrm{out}} =\left.\kappa_{+}\tilde{x}_{a}^{\mathrm{in}}\middle/\kappa_{-}\right.$
and
\begin{equation}
\tilde{p}_{a}^{\mathrm{out}} = \frac{g^{2}\chi_{m}\kappa
}{\kappa_{-}^{2}}\tilde{x}_{a}^{\mathrm{in}}
+\frac{\kappa_{+}}{\kappa_{-}}\tilde{p}_{a}^{\mathrm{in}}
-\frac{g\chi_{m}\sqrt{2\kappa\gamma_{m}}}{\kappa_{-}}\tilde{f}_{\mathrm{in}}.
\end{equation}
where $\kappa_{\pm}=\left.\kappa\middle/2\right.\pm i\omega$. Apparently, there is no
mechanical-mode information on the quadrature $\tilde{x}_{a}^{\mathrm{out}}$ and we
focus upon only the case where $\varphi=\left.\pi\middle/2\right.$, so that the total external
force can be expressed as:
\begin{align}
  F &= \frac{1}{F_{f}\left(\omega\right)}x_{a,\mathrm{out}}^{\pi/2}
  =\frac{1}{F_{f}\left(\omega\right)}\tilde{p}_{a}^{\mathrm{out}}
  =\tilde{f}_{\mathrm{ex}}+\tilde{f}_{\mathrm{add}} \nonumber\\
  &= \tilde{f}_{\mathrm{th}}+X_{a}\left(\omega\right)\tilde{x}_{a}^{\mathrm{in}}
  +P_{a}\left(\omega\right)\tilde{p}_{a}^{\mathrm{in}},
\end{align}
where $X_a\left(\omega\right)$, $P_a\left(\omega\right)$, and
$F_f\left(\omega\right)$ have been defined in Eq.~(\ref{key equations}). Then, we have obtained the induced force:
\begin{equation}
  f_{\mathrm{add}}\left(\omega\right)=\tilde{f}_{\mathrm{th}}
  +X_{a}\left(\omega\right)\tilde{x}_{a}^{\mathrm{in}}
  +P_{a}\left(\omega\right)\tilde{p}_{a}^{\mathrm{in}}.
\end{equation}
Moreover, we find that
$X_{a}^{*}\left(\omega\right)=X_{a}\left(-\omega\right)$,
$P_{a}^{*}\left(\omega\right)=P_{a}\left(-\omega\right)$.

The sensitivity of force measurement is commonly characterized
via the noise power spectral density ${S}_{\mathrm{F}}$,
which is given by~\cite{xu2014squeezing}:
\begin{equation}
  {S}_{\mathrm{F}}\left(\omega\right)
  = \int\mathrm{d}\omega'\left\langle f_{\mathrm{add}}\left(\omega\right)f_{\mathrm{add}}
  \left(\omega'\right)\right\rangle.
\end{equation}
Utilizing the correlation functions of the input vacuum
noise~\cite{xu2014squeezing}:
\begin{align}\label{x-p-fun}
\left\langle x_{a}^{\mathrm{in}}\left(\omega\right)
x_{a}^{\mathrm{in}}\left(\omega'\right)\right\rangle
&=\left\langle
p_{a}^{\mathrm{in}}\left(\omega\right)p_{a}^{\mathrm{in}}
\left(\omega'\right)\right\rangle
= \frac{1}{2}\delta\left(\omega+\omega'\right), \nonumber\\
\left\langle x_{a}^{\mathrm{in}}\left(\omega\right)
p_{a}^{\mathrm{in}}\left(\omega'\right)\right\rangle
&=-\left\langle p_{a}^{\mathrm{in}}\left(\omega\right)
x_{a}^{\mathrm{in}}\left(\omega'\right)\right\rangle =
\frac{i}{2}\delta\left(\omega+\omega'\right),
\end{align}
we obtain the symmetrized noise spectral density $S_{\mathrm{FF}}$ given in Eq.~(\ref{symmetrised noise power spectral density}).

\end{appendix}

%--------------------------------------------------------------------------------------%


\end{document}